# The solar chromosphere as induction disk and the inverse Joule-Thomson effect

Claudio Vita-Finzi
Dept of Earth Sciences, Natural History Museum
London SW7 5BD, UK (cvitafinzi@aol.com)

15 September 2016

*The connection between nuclear fusion in the Sun's core and solar irradiance is obscured among other things by uncertainty over the mechanism of coronal heating. Data for solar wind density and velocity, sunspot number, and EUV flux suggest that electromagnetic energy from the Sun's convection zone is converted by induction through the chromosphere into thermal energy. The helium and hydrogen mixture exhaled by the Sun is then heated by the inverse Joule-Thomson effect when it expands via the corona into space. The almost complete shutdown of the solar wind on 10-11 May 1999 demonstrated that its velocity is a more faithful indicator of solar activity than are sunspots as it reflects short-term variations in coronal heating rather than quasicyclical fluctuations in the Sun's magnetism. Its reconstruction from the cosmic ray flux using isotopes spanning over 800,000 yr should therefore benefit the analysis and long-term forecasting of Earth and space weather.*

Most of the mechanisms currently favoured for the rise in kinetic temperature from ~6,000 K at the Sun's photosphere to > 1-2 million K at the outer margins of the corona hinge either on magnetic reconnection or on MHD wave heating [1] but to judge from the literature neither is deemed to be wholly satisfactory. This letter eschews searching for periodicities by Fourier, wavelet and similar analyses to avoid the associated assumptions and data sacrifice, and it instead seeks major episodes within the various solar records taking advantage of the great improvement of recent years in the frequency of measurement for EUV and solar wind data. The critical zones discussed here are the Sun's convection zone, the chromosphere and the corona. The discussion is based mainly on photospheric imaging, measurements by the EVE instrument on the SDO spacecraft for the first six months of 2012, which allow energy transfer to be traced at different thermal levels within the corona, and a fall in solar wind speed [2] by 98 % in 10-12 May 1999.

Activity in the convection zone is reflected in the photospheric solar granulation, commonly viewed in three size categories, with cell diameters of ~ $10^6$ m, ~ 5 x $10^6$ m and ~3 x $10^7$ m (supergranules); in addition the existence of giant convection cells with diameters of ~2 x $10^8$ m has been inferred from the motion of the supergranules.[3] To these features may be added magnetic vortex structures which can be traced into the corona.[4] Atmospheric Imaging Assembly (AIA) observations on NASA's Solar Dynamics Observatory (SDO) satellite indicate EUV cyclones rooted in rotating network magnetic fields (RNFs), which are shown by the SDO's Helioseismic Imager (HMI) to be ubiquitous.[5] As they contribute 5.8 x $10^{22}$ Mx or 78% of the total unsigned magnetic flux, the RNFs emerge as prime candidates for conveying magnetic energy from the photosphere into the chromosphere.

Our proposal is for induction heating of the weakly ionised hydrogen-alpha of the chromosphere [6] by Foucault or eddy currents generated by rotating magnetic fields in the upper photosphere. Gas heating by induction was demonstrated many years ago in the laboratory by two NASA-supported studies. The first employed a DC arc jet and pre-ionised argon at low pressure but high velocity flowing through an RF induction heating duct.[7] In the second an induction plasma torch was successfully operated using hydrogen; the unit was ignited on pure argon but run on pure hydrogen [8] at one atmosphere at 60-160 kW. More recently, magnetic induction has been invoked [9] to account for the acceleration of the solar wind in polar coronal holes, while coronal heating by Joule dissipation of electric currents in the presence of neutral atoms, i.e. by ambipolar diffusion, has been shown to be very effective,[10] and induction heating has been modelled invoking a flat spiral work coil driven by a high frequency power inverter connected to an AC supply.[11] Spiral coils in series have been proposed in order to obtain a uniform magnetic field for near-field wireless power transfer. [12] The distinctive colour of the chromosphere, dominated by $H_\alpha$ lines (656.3 nm), indicates hydrogen ionisation: optically thin hydrogen plasma becomes 98 % ionised at an electron temperature of 20 x $10^3$ K, broadly consistent with the range estimated for the chromosphere from Skylab. The temperature decreases from the base of the chromosphere at about 6 x $10^3$ K to ~4 x $10^3$ K before increasing to ≥ 25 x $10^3$ K at a transition layer, a measure of the radial distance required for the Foucault effect to come fully into play.

The Extreme UV Variability Experiment (EVE) instrument on the SDO provides the opportunity to test this suggestion by tracing time variation in emission at a range of wavelengths (the EVE data [13] include EUV irradiance from 0.1 to 105 nm with a resolution of 0.1 nm, a temporal cadence of 10 s and an accuracy of 20 % ) for comparison with daily sunspot totals and with the solar wind. The 6-month period 1 Jan-30 June 2012 was selected for preliminary study because it lies near the midpoint on the rising limb of Solar Cycle 24 rather than at an extreme position in activity. Broad agreement was found [14, 15] between the fluctuations displayed by sunspot data for the visible disk (SSN), the disk integrated emission from the Sun at the radio wavelength of 10.7cm or 2800 MHz ($F_{10.7}$) which is a measure of *total* magnetic flux emerging through the photosphere, [16] and irradiance in watts $m^{-2}$ for the coronal lines for He II to Fe XVIII. The He II line is characteristic of the transition zone between the chromosphere and the corona, and data from rocket spectroheliograms [17] yield a height of $2820 \pm 400$ km and a log K of 4.9 for it. The other lines range in postulated formation temperature up to $10^{6.8}$ K. The turning points are in fair agreement (even if the details are not) and there is a general decrease in the amplitude of the fluctuations with formation temperature consistent with the HINODE/EIS finding that the amplitude of 3 min and 5 min oscillations propagating through the chromosphere and the transition decrease with increasing temperature within the lower corona. [18] In our scheme the chromosphere plays the role of the induction disks that are used to make non-ferrous cookware susceptible to heating by induction.

The broad synchroneity in the timing of peaks and troughs points to substantial radial rigidity in the corona, which some ascribe to continual magnetic reconnection but which is also consistent with heat transmission. A search for radial variations was pursued by referring to 1 minute EVE averages for Fe XII and Fe XIV on 20 January 2012 when there appeared to be a discrepancy in the timing of a clear peak amounting to as much as 2.5 hrs or 0.1 day (Fig. 1A). Being cooler than the Fe XX, the Fe XVI emission peaks on average 6 mins after the Fe XX and the GOES X-ray peak, and the time delay indicates the cooling rate of the post-flare coronal loops in the volume involved in both the impulsive and the gradual phases. For a long duration event (LDE: C3.2 flare) on 1 August 2010 the delay for the Fe XVI gradual phase peak was reportedly 101 minutes. In the situation depicted in Fig. 1A the hotter line peaked before the cooler by two hours. Care had been taken to rule out flare activity by selecting a period following the total decay of a substantial flare (active region 11402) which gave rise to M2.6 and M3.2 class solar flares and a full halo coronal mass ejection (CME): activity levels had apparently fallen to normal background levels as indicated by X-ray flux at 0.5-8.0 nm (Fig. 1B). In short, we have evidence of heat dissipation by a non-pulsatory process contrary to models which invoke large-scale reconnection or MHD waves.

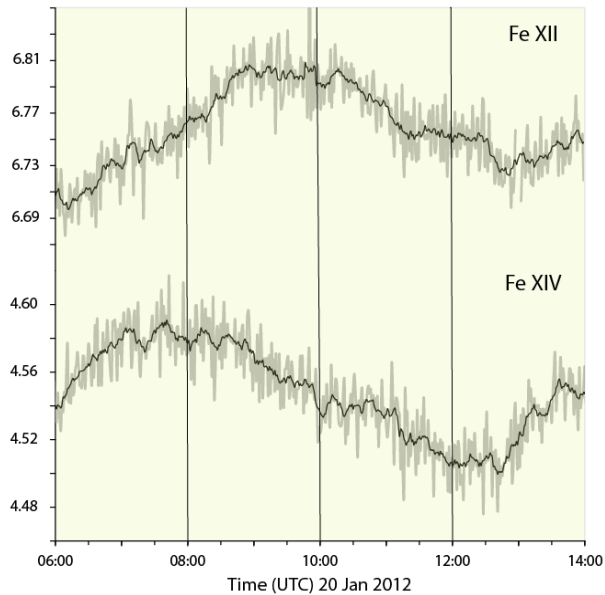

Fig 1A  Output of irradiance from Fe XII and Fe XIV coronal lines (plotted as 10 minute moving averages) on 20 January 2012. EVE/SDO ( data from  lasp.colorado.edu/  (courtesy of NASA).

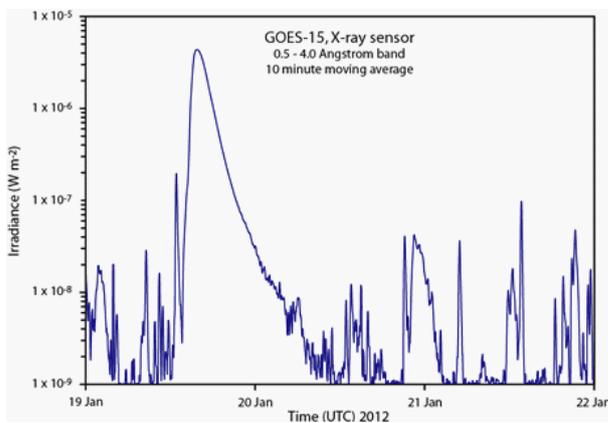

Fig 1B  Flare detected by X-ray sensor on GOES-15 (10 min. moving average) in January 2012. From  ngdc.noaa.gov/   Courtesy of NASA.

An unusual reduction in the solar wind density by 98% and in its speed by about a half (Fig. 2A) occurred [2] on 10-11 May 1999. The geomagnetic value at Earth Kp also fell to zero; [19] the sunspot record for the same period, however, shows no reduction, [20] and the values for $F_{10.7}$ [21] confirm a NASA report that there were no anomalies in the solar EUV flux as observed by the SOHO spacecraft. The prima facie case, therefore, is that fluctuations in the speed of the solar wind are largely independent of processes in the lower corona EUV represented by $F_{10.7}$ and the photosphere (sunspots) and are therefore a useful guide to energy transformation within the main body of the corona. Fig 2B shows that, for April-June 1999 (May 10 is day 131) the 275 MHz line, representative of the upper corona, departs violently from the trend for the lower corona (405-1755 MHz).

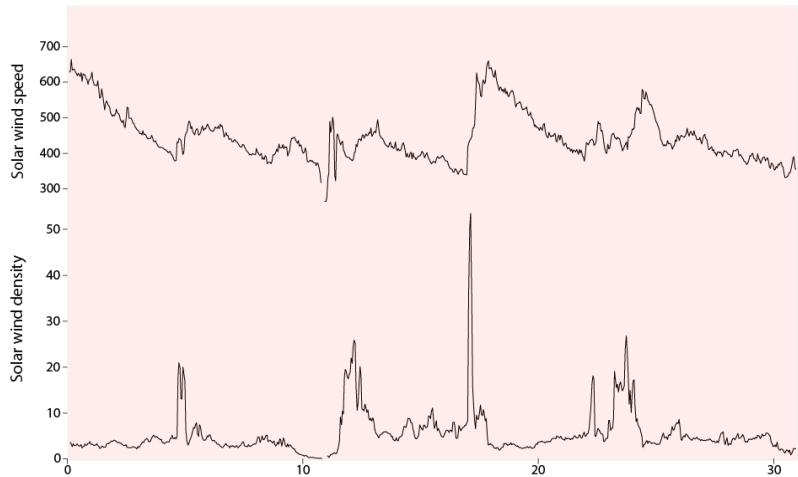

Fig 2A   Solar wind density (upper) and velocity (lower) for May 1999 from srl.caltech.edu/cgi-bin/dib/rundibviewswel2/ACE/ASC/DATA/level2/

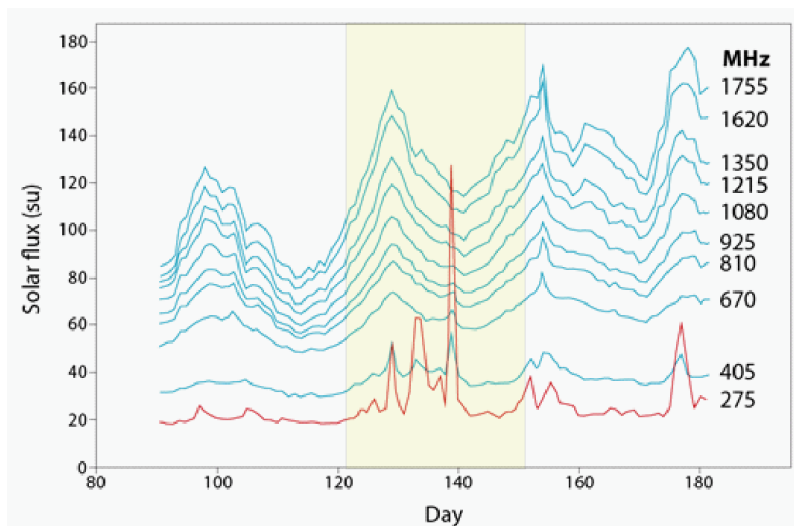

Fig 2B   Extract from solar radio observations for 1999, Astronomical Observatory of the Jagiellonian University, Cracow, at oa.uj.edu.pl/index.en.html. Shading indicates the month of May.

On the whole experimental work in this field is focused on plasmas in tokamaks, where temperatures of $1.5 \times 10^8$ K are obtained through ohmic heating derived from an induced current supplemented by RF heating. A major hurdle in such laboratory studies is the need to confine the plasma, usually attained magnetically, but the solar environment may bring this about gravitationally (escape velocity for $H_2$ on the Sun is 6.2 x $10^5$ m $s^{-1}$). A clue to the route by which heat is transmitted through the corona to the solar wind is provided by the predominance of H and He in the Sun's composition. These are two of the three gases (the third is neon) with such a low inversion temperature that above ambient temperature and under isoenthalpic conditions their response to the Joule-Thomson effect is heating rather than cooling.[22] In the absence of experimental data for coronal temperatures we must rely on extrapolation of inversion curves,[23] but to advance the discussion we note that at very high temperatures the Joule-Thomson coefficient $\mu_{JT}$ may be represented by – $b/C_p$,[24] where $b$ is of course one of the two constants that distinguish the van der Waals equation from the Real Gas Law and that vary according to the gas at issue, and $C_p$ is the heat capacity. The negative result is consistent with the temperature increase obtained experimentally and by calculation. The question now is how far it applies at the chromosphere and the corona. The radial elevation of the temperature maximum above the photosphere [25] provides a rough measure of the limits of J-T heating and it illustrates how astronomical observation can supplant the laboratory for evaluating thermochemical processes at extreme settings.

The solar wind emerges as the preferred indicator of solar activity. Sunspot data are compromised by their indirect relation to the Sun's irradiance: the rotation of active areas explains no more than 42% of its variation.[26] Moreover the secure sunspot record spans at most four centuries and says little about such matters as solar fluctuations during the Maunder and other sunspot minima. It remains to be seen whether the cosmogenic isotope record, which already spans 800,000 years,[27] can provide the requisite link to the history of the solar wind on the grounds that it represents its modulating effect on the flux of galactic cosmic rays.[28]